\documentclass{ws-procs10x7}
\usepackage{balance}

\newcolumntype{d}[1]{D{.}{.}{#1}}

\def\Journal#1#2#3#4{{\it #1} {\bf #2}, #3 (#4)}

\makeindex
\begin{document}

\title{Prompt Photons and Particle Momentum Distributions at HERA}

\author{N. Gogitidze}

\address{DESY, Notkestrasse 85, 22607 Hamburg, Germany\\
E-mail: nellyg@mail.desy.de}


\twocolumn[\maketitle\abstract{Recent results, obtained by the H1 and ZEUS
collaborations, are presented on differential cross sections,
for inclusive prompt photon production in DIS and for photoproduction 
of prompt photons accompanied by a hadronic jet. 
Also presented are cross sections of normalised scaled 
momentum distribution of charged final state hadrons, measured  by H1 
in DIS $ep$ collisions at high $Q^2$ in the Breit frame of reference. }
]

\section{Prompt photon production}
\subsection{Introduction}\label{subsec:intro1}

Isolated high transverse energy  photons in the final state 
are a powerful tool for detailed studies of the Quantum Chromodynamics (QCD)
in hard interaction processes and of the hadronic structure of the 
incoming particles.

The photons are called ``prompt'' if they are directly coupled to the 
interacting quarks, instead of being produced as hadronic decay products.

In contrast to jet measurements, where the partonic structure is
obscured by the non-perturbative hadronisation process, prompt photons
at large transverse energy $E_{T}^{\gamma}$ can be directly related to
the partonic event structure. Furthermore, the experimental uncertainties
connected with the energy determination of an electromagnetic shower
initiated by a photon are smaller compared to the measurement of a hadron jet.
However, the cross section for prompt photon production is small
and the identification of photons in the detector is not trivial.

Preliminary results for two analyses are presented here: 
An H1 study of inclusive prompt photons in deep inelastic scattering
(DIS)\cite{pph1}, 
and a ZEUS study of prompt photons with an accompanying jet in
photoproduction\cite{ppzeus}. 

\subsection{Prompt~Photon~identification}\label{subsec:ppident}

The main experimental difficulty is the separation of the prompt photons
from hadronic background, in particular from signals due to $\pi^{0}$ mesons.

In the H1 analysis photons are identified in the liquid argon calorimeter
by a compact electromagnetic cluster with no track pointing to it.
The photon transverse energy $E^{\gamma}_{T}$ and pseudorapidity 
$\eta^{\gamma}$ are restricted to 3~GeV~$<~E^{\gamma}_{T}~<$~10~GeV and 
$-1.2~<~\eta^{\gamma}~<~1.8$.  In order to compare with perturbative QCD
(pQCD) calculations,
the photon isolation requirement is defined in an infrared-safe way,
using:  $z = E^{\gamma} / E^{photonjet} > 0.9$,
i.e. $z$ is the ratio of the photon energy to the energy
of the jet, which contains the photon.
The photon signal is extracted by a shower shape analysis which uses six
discriminating shower shape functions in a likelihood method.

In the ZEUS analysis photons are identified using the barrel preshower 
detector (BPRE). The BPRE prompt photon signal is determined using the
conversion probability in the detector, known from a study of DVCS 
data\cite{dvcs}.
 
The photon kinematic range is restricted to 
$5~< E^{\gamma}_{T} < 16$~GeV and
$-0.7~<~\eta^{\gamma}~<~1.1$, where positive $\eta^{\gamma}$ corresponds
to the proton beam direction. The photon isolation criteria are similar
to the ones used in the H1 analysis. 
Hadronic jets were selected in the kinematic range 
$6<E^{jet}_{T}<17$~GeV and $-1.6<\eta^{jet}<2.4$. 
    
\subsection{Results}\label{subsec:ppres}

Differential cross sections for the production of isolated photons in DIS
measured by H1 are shown in Fig. 1, as function of $E^{\gamma}_{T}$ and
$\eta^{\gamma}$. A new LO($\alpha^{3}$) 
calculation\cite{lo1} gives a good description, although it lies slightly
below the data.
At large pseudorapidities the dominant contribution is radiation off
the quark line (QQ), whereas in the backward region the radiation off the
electron line (LL) dominates the cross section. 
\begin{figure}
\hspace*{0.3cm}
\includegraphics[width=0.45\textwidth]{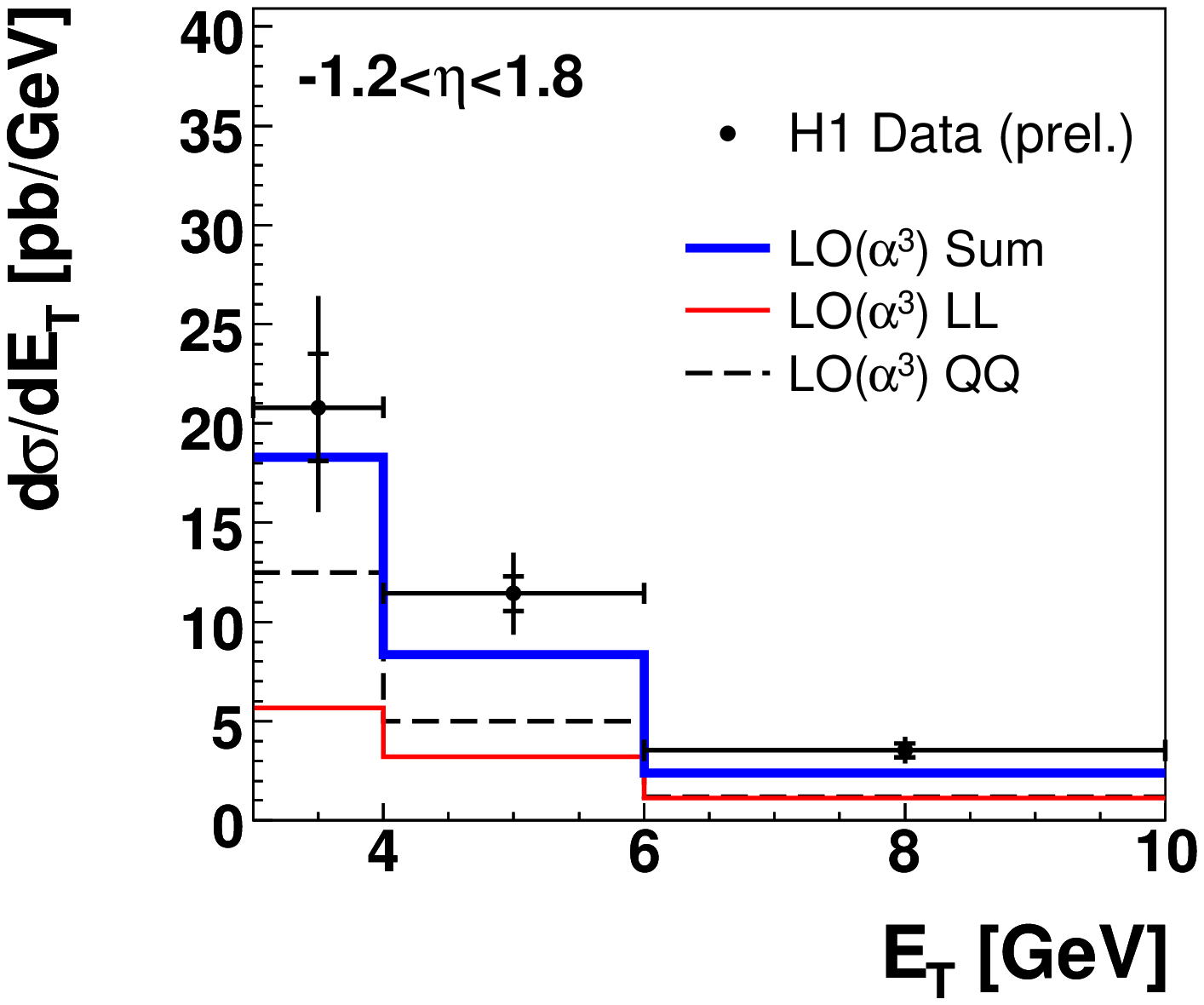}
\hspace*{0.3cm}
\includegraphics[width=0.45\textwidth]{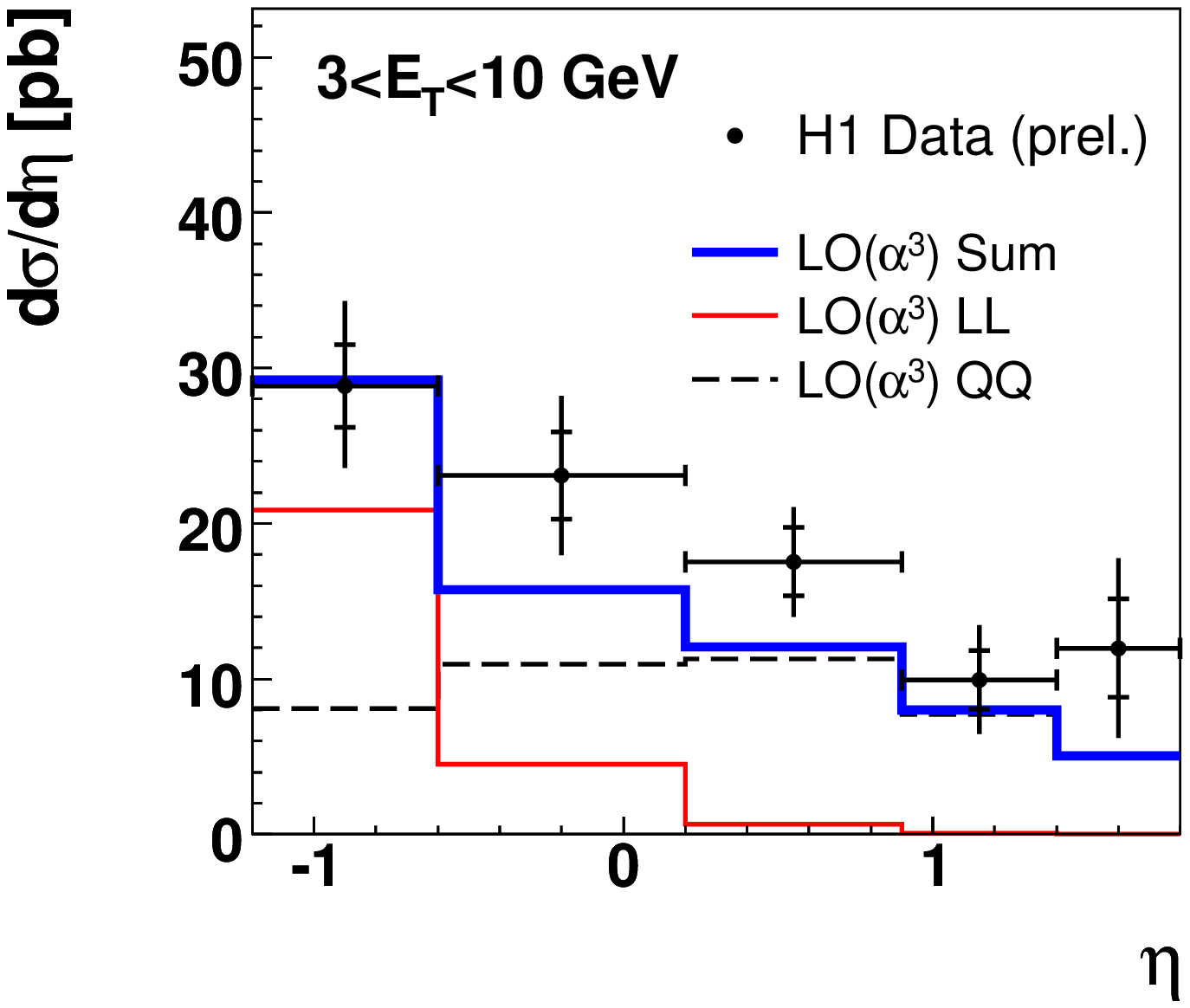}
\caption{ Prompt photon differential cross sections $d\sigma/dE^{\gamma}_{T}$ 
for $-1.2< \eta^{\gamma} <1.8$  and  $d\sigma/d\eta^{\gamma}$ for 
3~GeV$< E^{\gamma}_{T} <$ 10~GeV, for photon virtualities 
$Q^2 > $ 4~GeV$^{2}$ and
$y_{e} >$ 0.05. Curves show an LO calculation with LL and QQ giving the
contribution of radiation off the electron and the quark line respectively.
As the interference is very small it is not shown, but included in the sum.}
\label{fig1}
\end{figure}

A comparison with predictions of the PYTHIA and HERWIG 
generators (radiation off the quark) plus photon radiation off the electron
is also made\cite{pph1}.
Both generators describe the shape in $E^{\gamma}_{T}$ well, but are 
\hyphenation{signi-fi-cantly}
lower in the absolute scale (factor 2.3 for PYTHIA and 2.6 for HERWIG). 
The $\eta^{\gamma}$ distribution is better described by PYTHIA. 

The ZEUS differential cross sections as functions of $E^{\gamma}_{T}$ and 
$\eta^{\gamma}$  for the prompt photons are shown in Fig.~2. 
Two next-to-\hyphenation{lead-ing} order (NLO) pQCD predictions are 
compared to the data. 
In both calculations several photon and proton pdf's, and several 
fragmentation functions are used. 
The FGH (Fontannaz, Guillet and Heinrich)\cite{fgh} calculation contains 
additional
higher order corrections to the resolved photon process. Like the KZ
(Krawczyk and Zembrzuski)\cite{kz} prediction, it describes the data 
rather well.
However, they both underestimate the observed cross section at low 
$E^{\gamma}_{T}$ and in the backward region. 
The difference between the data and the
NLO QCD calculations is mainly observed in the $x_{\gamma}^{obs}~ <$~ 0.75
region (not shown), which is sensitive to the resolved photon contribution.
A comparison with the prediction of Lipatov and Zotov\cite{lz} (LZ), which 
is based on  
$k_{T}$-factorisation, corrected for hadronisation effects, is also shown.
The LZ prediction gives the best description of the $E^{\gamma}_{T}$ and 
$\eta^{\gamma}$ cross sections. In particular, it describes the 
lowest $E^{\gamma}_{T}$ region better than the KZ and FGH NLO predictions.
PYTHIA and HERWIG do not rise as steeply
at low $E^{\gamma}_{T}$ as do the data and underestimate the measured
cross section. 
\begin{figure}
\includegraphics[width=0.5\textwidth]{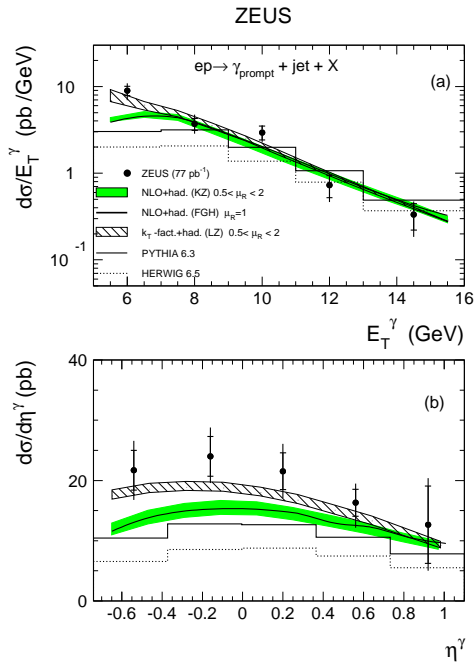}
\caption{ The differential cross section for the prompt photon events with an
accompanying jet as functions of $E^{\gamma}_{T}$ and $\eta^{\gamma}$
compared to theoretical QCD calculations (including hadronisation corrections).
The shaded bands correspond to the uncertainty in the renormalisation scale
which was changed by factors 0.5 and 2.}
\label{fig2}
\end{figure} 

Since the largest difference between the NLO calculations and the data is 
observed in the region of low $E^{\gamma}_{T}$ and low $\eta^{\gamma}$,
the level of agreement with NLO QCD was verified by increasing the minimum 
transverse energy of prompt photons from 5 to 7 GeV. In this case
hadronisation corrections are expected to be smaller. As shown in Fig.~3
with $E^{\gamma}_{T}~>~7$~GeV the NLO QCD and the LZ
predictions all agree well with the data. 
The PYTHIA model then also agrees well, while HERWIG is still below the data. 
\begin{figure}
\includegraphics[width=0.5\textwidth]{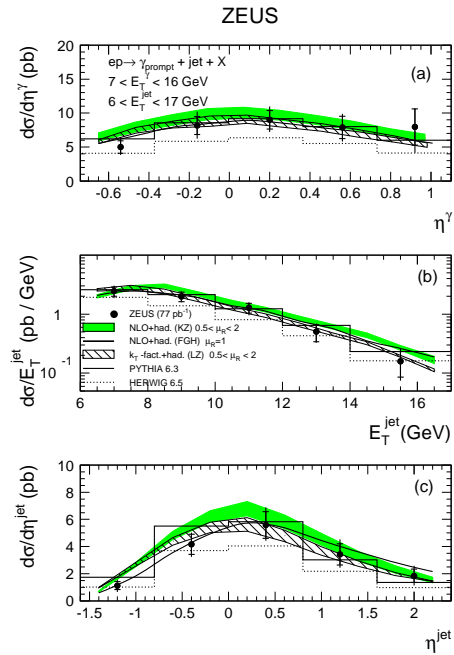}
\caption{ The differential cross section for the $\gamma$ + jet events as 
function of: a) $\eta^{\gamma}$, b) E$^{jet}_{T}$ and c) $\eta^{jet}$ compared
to QCD calculations (with hadronisation corrections) and Monte Carlo 
models. The cut on $E^{\gamma}_{T}$ is increased to 7 GeV. }
\label{fig3}
\end{figure}
\section{Charged Particle Momentum distributions}

In an H1 analysis\cite{pmdh1} the process of parton fragmentation and 
hadronisation is studied using inclusive charged particle spectra in 
DIS. 
In the current region of the Breit frame a comparison with one 
hemisphere of $e^{+}e^{-}$ annihilation offers a direct possibility 
to test quark fragmentation universality. 

The energy scale for the current region, set by the virtual
photon, is given by $Q/2$ and, for purpose of comparison, is taken to be
equivalent to one half of the $e^{+}e^{-}$ c.m. energy  $E^{*}/2$.

In the Breit frame the scaled momentum variable $x_p$ is defined to be
$2p^{\pm}_{h} / Q$, where $p^{\pm}_{h}$ is the momentum of a charged particle. 
In $e^{+}e^{-}$ annihilation events the equivalent variable is 
$2p^{\pm}_{h} / E^{*}$.

The use of much higher statistics now available at
high $Q$ as  compared to previous studies\cite{st1,st2}, as well as an improved
understanding of the H1 detector and associated systematics, provide a much
improved measurement of the scaled momentum spectra. Results are now available
up to $<Q> \sim$ 100 GeV, close to the LEP-1 c.m. energy, and in the
full range of $x_p$ (0 $< x_p <$ 1).

In Fig. 4 the inclusive, event normalised, charged particle scaled momentum
spectrum is shown as a function of $Q$ for nine different bins of $x_p$.
Also shown is a comparison to results from  $e^{+}e^{-}$ annihilation 
data (see references in ref. \refcite{pmdh1}). 
As seen, the $ep$ and $e^{+}e^{-}$ data
are in excellent agreement, which supports the concept of 
quark fragmentation universality.  

Moving from low to high $Q$ the $x_p$ spectra become softer, i.e. there is a
dramatic increase in the number of 
hadrons with a small share of the initial parton's momentum and a decrease
of those hadrons with a large share.
These scaling violations (parton splitting in QCD) are compatible
to the scaling violations observed for the DIS structure functions. 

In the RAPGAP simulation\cite{rapgap}, also shown in Fig. 4, the 
Parton Shower model is implemented.  
It describes the fragmentation process as the splitting
of parent partons into two daughters (e.g. $q \rightarrow qg, 
q \rightarrow qq, g \rightarrow qq$), the splitting continues with daughters
going on to form parents. The evolution of the parton shower is based on
leading $\log{Q^2}$ DGLAP splitting functions. RAPGAP gives a very good 
description of the $ep$ scaled momentum spectra over the whole range of $x_p$. 
\begin{figure}
\includegraphics[width=0.5\textwidth]{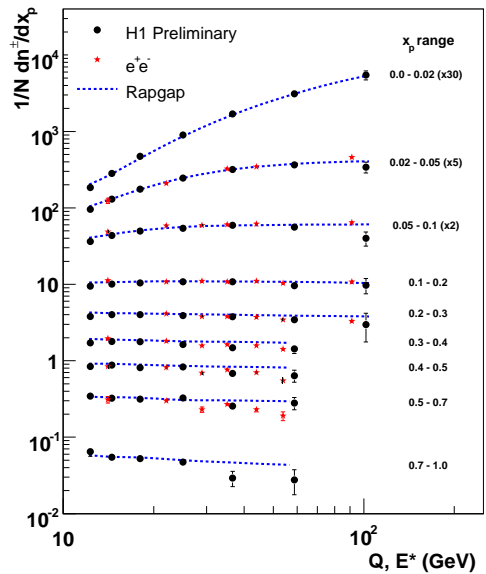}
\caption{ H1 data for the event normalised inclusive scaled momentum
spectrum as a function of $Q$ for nine different $x_p$ regions. 
Also shown are data from various $e^{+}e^{-}$ experiments 
(taking $Q = E^{*}$). The DIS data are compared with the RAPGAP generator.}
\label{fig4}
\end{figure}

\end{document}